\documentclass[11pt]{article}
\usepackage[a4paper,margin=1in,top=25mm]{geometry}
\usepackage{amsmath,amssymb,amsthm}
\usepackage{graphicx}
\usepackage{hyperref}
\usepackage{authblk}

\title{Coupled Topological Interface States and Phonon Molecules in GaAs/AlAs Superlattices}
\author[1]{S. Sandeep}
\author[1]{O. Colmegna}
\author[1]{C. Xiang}
\author[1]{E. R. Cardozo de Oliveira}
\author[1]{K. Papatryfonos}
\author[1]{M. Morassi}
\author[1]{A. Lemaitre}
\author[1]{N. D. Lanzillotti-Kimura\thanks{Corresponding author: daniel.kimura@cnrs.fr}}

\affil[1]{Université Paris-Saclay, C.N.R.S., Centre de Nanosciences et de Nanotechnologies (C2N),  10 Boulevard Thomas Gobert, 91120 Palaiseau, France}

\begin{document}

\maketitle

\begin{abstract}
Topological interface states in one-dimensional superlattices provide spatially localized phonon modes protected by the topology of the underlying band structure. In GaAs/AlAs distributed Bragg reflectors (DBRs), such states can be engineered through band inversion between superlattices with opposite Zak phases within the Su–Schrieffer–Heeger (SSH) framework. Here, we demonstrate topological phonon molecules and extended chains formed by coupled nanophononic interface states. By concatenating three superlattices with alternating topology, we realize two coupled interface states that hybridize into symmetric and antisymmetric modes, whose splitting can be tuned over tens of gigahertz by varying the reflectivity of the central DBR. Extending this concept, we engineer chains of up to $N=6$ coupled interface states that form narrow topological minibands while remaining strongly localized at the interfaces. We experimentally observe these coupled states in molecular-beam-epitaxy-grown GaAs/AlAs heterostructures using time-domain pump–probe transient reflectivity measurements, and reproduce their behavior using transfer-matrix calculations and a simple analytical model for the mode splitting. These results establish topological interface states as a robust platform for engineering coupled phononic systems and tunable nanophononic architectures in the GHz regime.
\end{abstract}

\section{Introduction} \label{introduction}

A one-dimensional periodic lattice containing two elements per unit cell can be described within the Su–Schrieffer–Heeger tight‑binding model, a paradigmatic framework for understanding topological phases and localized excitations in condensed matter \cite{su_solitons_1979, asboth_short_2016}. This model, originally introduced to explain
the electronic properties of polyacetylene \cite{su_solitons_1979}, has since inspired the realization of topological states
for a broad variety of excitations, including photons \cite{ozawa_topological_2019, hauff_chiral_2022, hafezi_imaging_2013, xiao_surface_2014, lu_topological_2014, khanikaev_two-dimensional_2017}, phonons \cite{rodriguez_2023,esmann_topological_2018, arregui_coherent_2019,PhysRevB.97.020102, susstrunk_observation_2015, prodan_topological_2009}, mechanical vibrations \cite{doster_observing_2022, kettler_inducing_2021, ma_topological_2019,xiao_geometric_2015,zhao_topological_2018,zheng_observation_2019,zhang_topological_2018,he_acoustic_2016,huber_topological_2016, leeElasticTopologicalInterface2022}, polaritons \cite{st-jean_measuring_2021, pernet_gap_2022, chafatinos_polariton-driven_2020, kuznetsov_electrically_2021-1, klembt_exciton-polariton_2018}, plasmons \cite{song_plasmonic_2021},
and magnons \cite{malz_topological_2019}. In multilayered optical and acoustic structures, distributed Bragg reflectors exhibit high‑reflectivity bands associated with bandgaps in the underlying dispersion relation. The Bloch modes at the edges of these bandgaps possess well-defined spatial symmetries, and concatenating two DBRs whose band-edge modes have inverted symmetries around a common gap gives rise to a topological interface state within that gap \cite{ esmann_topological_2018, esmann_topological_2018_2, esmann_topological_2018_1, rodriguez_2023}. The existence and robustness of this interface state can be understood in
terms of the Zak phase—the one‑dimensional analogue of the Berry phase—which changes by $\pi$ across a band inversion.  

Over the past two decades, the field of nanophononics \cite{priya_perspectives_2023,Chushuang_2024, ortiz_fiber-integrated_2020}, concerned with the control of acoustic nanowaves in solids, has developed a comprehensive toolbox of phonon mirrors, filters, and resonant cavities based on acoustic impedance engineering \cite{esmann_topological_2018,ortiz_phonon_2019}. In contrast to optical or electronic systems, where the relevant dispersion is typically nonlinear \cite{parappurath_direct_2020,lateral_Bragg_Papatry,Wang_2018}, acoustic phonons exhibit a nearly linear dispersion \cite{delsing_2019_2019}, thereby enabling studies extending over a broad frequency range \cite{rodriguez_2023,arregui_coherent_2019}. Moreover, the relatively slow speed of sound and long coherence lengths in these structures allow one to engineer effectively quasi‑infinite lattices on the micron scale, while standard ultrafast optical techniques such as pump–probe spectroscopy and Brillouin scattering can provide direct access to their spectral and temporal response \cite{arregui_coherent_2019}. The control and manipulation of acoustic phonons, as well as their interactions, can have significant technological implications for thermal transport and heat management~\cite{torres_giant_2019,li_remarkably_2021,bencivenga_nanoscale_2019}, quantum applications~\cite{chu_creation_2018,buhler_-chip_2022,meng_measurement-based_2022}, and information exchange~\cite{mahboob_phonon-cavity_2012,stiller_coherently_2020}.

A key component to achieve phonon confinement and control is the distributed Bragg reflector. DBRs find a wide range of applications in both phononics and photonics, owing to their high reflectivities, which allow the formation of high-Q cavities, as well as their excellent material quality and reproducibility resulting from epitaxial growth by molecular beam epitaxy (MBE)\cite{papatryfonos_2025}. In particular, GaAs/AlAs DBRs can confine photons and high-frequency acoustic phonons in a single structure thanks to the comparable refractive index and acoustic impedance contrasts between GaAs and AlAs, a situation often dubbed “double magic coincidence” in the literature \cite{ortiz_topological_2021,fainsteinStrongOpticalMechanicalCoupling2013}. 

In this context, topological concepts have recently been brought into nanophononics. Theoretical and experimental works have shown that GaAs/AlAs DBRs can be tailored to exhibit band inversion in a chosen acoustic gap by adjusting the unit cell thickness ratio between the two constituent materials while preserving the total acoustic period ~\cite{esmann_topological_2018}. The concatenation of two such band‑inverted superlattices generates a symmetry‑protected nanophononic interface state that is pinned to the center of the bandgap and robust against thickness disorder that does not alter the underlying Zak phases~\cite{arregui_coherent_2019}. These ideas have been used to design and observe topological acoustic interface modes in GaAs/AlAs structures via Raman spectroscopy and time‑resolved pump–probe experiments ~\cite{arregui_coherent_2019}, to realize interface states simultaneously for light and sound through coincident optical and acoustic band inversion ~\cite{ortiz_topological_2021}, and to access higher‑order acoustic bandgaps to generate topological states across a broad frequency range ~\cite{rodriguez_2023, papatryfonos_SPIE_2024}. Beyond their fundamental interest, such topological nanophononic resonators may offer a robust platform for enhanced optomechanical interactions and hybrid quantum devices. Topological interface states have been shown to exhibit exceptional robustness \cite{rodriguez_2023,ortiz_topological_2021,esmann_brillouin_2019} against compositional fluctuations of the order typically encountered due to material intermixing or inhomogeneities during MBE growth~\cite{RefInd}, which is a highly desirable feature for transport and data communication applications~\cite{mathew_synthetic_2020,arora_direct_2021}.

Parallel to these developments, phonon engineering in multilayered structures has explored phonon molecules and more complex nanomechanical potentials in the conventional Fabry–Pérot picture ~\cite{lanzillotti-kimura_phonon_2007, ortiz_phonon_2019}. In early work, double and multiple acoustic cavities coupled through DBRs were theoretically proposed as phononic analogues of diatomic molecules, exhibiting symmetric and antisymmetric normal modes whose splitting and lifetimes can be tuned by the reflectivity of the central and external mirrors ~\cite{lanzillotti-kimura_phonon_2007}.
These concepts were later generalized to adiabatically modulated superlattices implementing effective one‑dimensional potentials such as parabolic wells, Morse potentials, and double wells for acoustic phonons ~\cite{ortiz_phonon_2019}. These structures can support minibands and Wannier–Stark ladders of phonon states and can mimic Bloch oscillations, opening an avenue to simulate quantum‑mechanical potentials and transport using coherent acoustic phonons ~\cite{lanzillotti-kimura_bloch_2010}.

Coupling between topological interface states has been studied in detail in the optical domain. Photonic crystals composed of dielectric or metal–dielectric multilayers have been used to realize pairs of topological Tamm or Zak phase interface states; when two such states are brought into proximity by concatenating three photonic segments with alternating topological phases, they hybridize into molecular‑like symmetric and antisymmetric modes with a characteristic mode splitting~\cite{schmidt_coupled_2021}. This splitting can be tuned by varying the thickness of the central segment and manifests in anticrossings and polarization‑dependent resonances in reflectivity spectra ~\cite{sharma_2023}. Yet, despite the rapid progress on individual nanophononic topological states ~\cite{rodriguez_2023, ortiz_topological_2021, esmann_topological_2018}, a unified framework where topologically protected nanophononic interface modes play the role of the “atoms” forming phonon molecules and extended chains, has remained largely unexplored.

In this work, we bridge these two perspectives by introducing topological nanophononic molecules and multi‑interface lattices built from SSH‑type interface states in GaAs/AlAs superlattices. We first revisit the design of a single topological nanophononic interface mode in the acoustic bandgap around $300~\mathrm{GHz}$, implemented by concatenating two DBRs with band-inverted centrosymmetric GaAs/AlAs unit cells centered on the AlAs layer. This structure provides the fundamental building block for our devices. We experimentally realize structures supporting topological phonon-molecule modes in a monolithic MBE-grown GaAs/AlAs heterostructure, by concatenating three superlattices with alternating Zak phases, yielding two symmetry-protected interface modes at the same frequency that couple through the central DBR. The hybridization of these modes results in a mode splitting between symmetric and antisymmetric combinations, whose magnitude we control by varying the number of periods—and hence the reflectivity—of the central mirror. We show that the splitting follows a simple expression in terms of an effective acoustic velocity of the structure, the central‑mirror reflectivity, and an effective coupling length, providing an intuitive design rule for the coupling strength. Importantly, the resulting hybrid modes retain the interface localization associated with the underlying topological band inversion and show enhanced robustness against thickness-ratio perturbations compared with conventional Fabry–Pérot phonon molecules.

We further generalize this strategy to engineer devices hosting multiple coupled interface states by concatenating extended sequences of topologically distinct superlattices. This yields narrow topological minibands formed by  $N$ = 6 coupled interface modes that remain spatially pinned to their respective interfaces, with tunable bandwidth controlled by the separation between successive interfaces ~\cite{lanzillotti-kimura_phonon_2007}. 

Building on established schemes for coherent phonon generation and detection in topological nanocavities ~\cite{ortiz_topological_2021} using time‑domain pump–probe transient reflectivity, we resolve the $300~\mathrm{GHz}$ interface resonance as well as its splitting into chain modes. Only one mode is observed in the molecular case; the other is symmetry-forbidden, in agreement with transfer‑matrix simulations.  In contrast to previous studies focusing on isolated interface states, our approach enables controlled coupling of multiple topological phononic modes, forming molecule- and chain-like architectures with tunable interactions.

The paper is organized as follows. In Sec. II, we review the formation of topological nanophononic interface states by band inversion in GaAs/AlAs superlattices. In Sec. III, we introduce the concept of topological phonon molecules, analyze the coupling between interface states and the resulting mode splitting, and describe their experimental realization and observation. Section IV addresses multi-interface structures supporting chains of coupled interface states and their experimental observation. We conclude in Sec. V with a discussion of future directions of the present work and prospects for nanophononic devices.

\section{Topological phononic interface states by band inversion} \label{interface states}

\begin{figure}[ht]
    \centering
S    \includegraphics[width=0.7\columnwidth]{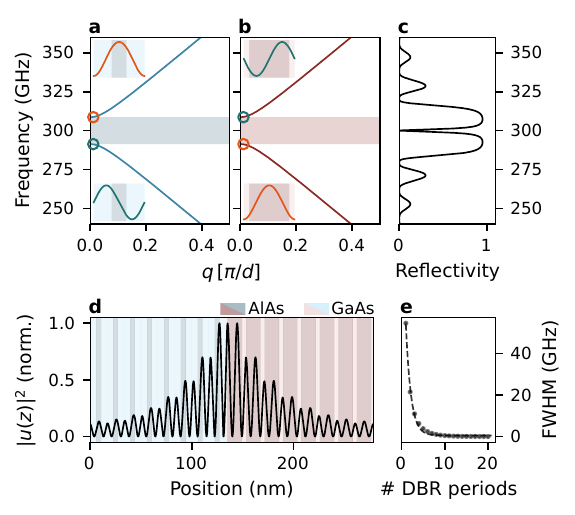}
    \caption{Principle of a phononic interface state formed by band inversion. 
(a) Acoustic dispersion of an infinite GaAs/AlAs superlattice with centrosymmetric unit cell $(3\lambda_{\mathrm{GaAs}}/8,\ \lambda_{\mathrm{AlAs}}/4,\ 3\lambda_{\mathrm{GaAs}}/8)$, designed for $300~\mathrm{GHz}$, showing a band gap at the Brillouin-zone center $(q=0)$. Insets: displacement profiles of the band-edge modes, exhibiting symmetric and antisymmetric patterns with respect to the unit-cell center (green and orange curves).(b) Dispersion of the band-inverted unit cell $(\lambda_{\mathrm{GaAs}}/8,\ 3\lambda_{\mathrm{AlAs}}/4,\ \lambda_{\mathrm{GaAs}}/8)$, which preserves the dispersion but reverses the band-edge symmetries.(c) Acoustic reflectivity of a finite structure formed by concatenating the two superlattices shown in (a) and (b), each consisting of 8 periods.  A localized resonance appears inside the common band gap ($\sim300~\mathrm{GHz}$), indicating the formation of a confined interface mode. (d) Spatial profile of the squared displacement $|u(z)|^2$ at the interface-mode frequency, localized at the DBR junction and decaying evanescently into the mirrors. (e) Full width at half maximum (FWHM) of the interface resonance as a function of the number of DBR periods. }
    \label{fig-interface}
\end{figure}

The infinite periodic arrangement of two materials with contrasting acoustic impedances, defined as the product of sound velocity and mass density, can be described by a folded Brillouin zone in the acoustic dispersion relation, where band gaps emerge at both the zone center and the zone edge. The opening and closing of these gaps depend on the relative thickness ratio between the constituent layers ~\cite{esmann_topological_2018_1, esmann_topological_2018_2}. In the present design, we consider centrosymmetric GaAs/AlAs unit cells centered on the AlAs layer. For a structure targeting an acoustic band gap around $300~\mathrm{GHz}$, the unit cell is composed of the layer sequence $(3\lambda_{\mathrm{GaAs}}/8,\ \lambda_{\mathrm{AlAs}}/4,\ 3\lambda_{\mathrm{GaAs}}/8)$, where $\lambda$ denotes the acoustic wavelength at the design frequency. This configuration opens a band gap at the Brillouin-zone center ($q=0$), as shown in Fig.~\ref{fig-interface}(a). The band-inverted counterpart (Fig.~\ref{fig-interface}(b)) is obtained by exchanging the thickness distribution between the two materials, leading to the unit cell $(\lambda_{\mathrm{GaAs}}/8,\ 3\lambda_{\mathrm{AlAs}}/4,\ \lambda_{\mathrm{GaAs}}/8)$. While both superlattices exhibit identical dispersion relations, the spatial symmetries of the band-edge modes are inverted. As a result, the two superlattices belong to distinct topological families.

In finite structures, the band gaps correspond to frequency regions of high acoustic reflectivity, characteristic of distributed Bragg reflectors. Band inversion arises when two DBRs that (i) share a common band gap and (ii) exhibit opposite spatial mode symmetries are concatenated (Fig.~\ref{fig-interface}(a,b)). When these conditions are satisfied, a topological interface mode forms at the junction between the two DBRs. This mode appears as a dip in the high-reflectivity region, as shown in Fig.~\ref{fig-interface}(c) for a structure composed of 8 periods in each superlattice. The spatial profile of the squared displacement $|u (z)|^2$ of the mode at $300~\mathrm{GHz}$ is shown in Fig.~\ref{fig-interface}(d), where the displacement reaches its maximum at the interface.

The linewidth of the interface mode decreases as the number of DBR periods in each superlattice increases, as shown in Fig.~\ref{fig-interface}(e). This trend arises from the enhanced acoustic confinement provided by thicker Bragg mirrors, which increases the fraction of energy stored near the interface and reduces leakage into the surrounding layers. The mechanism is analogous to the linewidth narrowing observed in conventional Fabry–Pérot resonators when the mirror reflectivity is increased ~\cite{lanzillotti-kimura_phonon_2007}.

Nanophononic interface states achieved by band inversion in the SSH framework have been previously studied, both numerically and experimentally ~\cite{esmann_topological_2018_1,ortiz_topological_2021,rodriguez_2023,arregui_coherent_2019}. These works have shown that the interface mode remains robust against certain variations in the thickness of the layers constituting the DBRs, characteristic of topological protection. In the present context, this interface state serves as the fundamental building block for the engineering of more complex structures explored in the following sections.

%\section{Engineering topological %phononic molecules} \label{topological %molecule}

\section{Coupled interface states: Topological phonon molecules} \label{topological molecule}
\begin{figure}[ht]
    \centering
    \includegraphics[width=0.6\columnwidth]{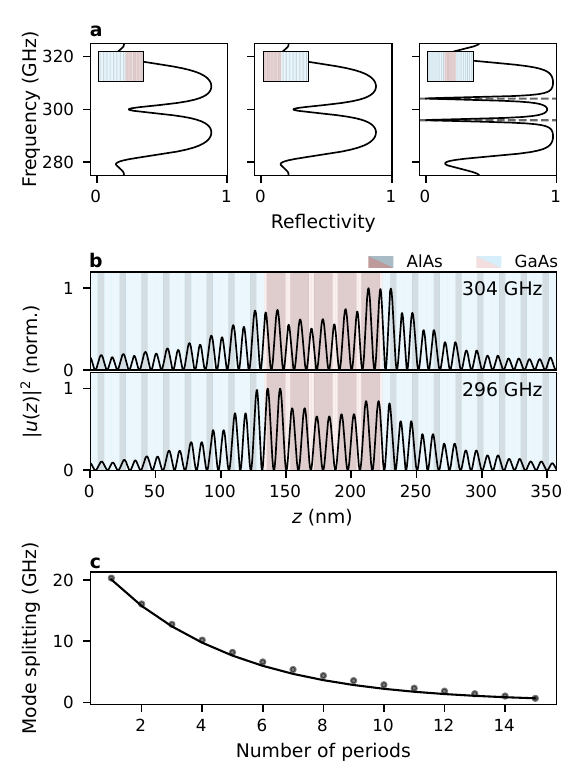}
    \caption{Topological phonon molecules formed by coupling two interface states.
(a) Acoustic reflectivity of structures illustrating the formation of coupled interface states: a single interface obtained by concatenating two band-inverted superlattices, the reverse concatenation producing an equivalent interface state, and a three-segment structure containing two interfaces.
(b) Spatial profiles of the squared displacement of the two hybridized interface modes.
(c) Mode splitting as a function of the number of periods in the central DBR. Increasing the mirror thickness reduces the coupling and therefore decreases the splitting.}
    \label{fig-molecule-design}
\end{figure}

A topological phonon molecule can be understood as a pair of coupled topological bound states within an effective SSH-like framework, analogous to a double-cavity structure~\cite{lanzillotti-kimura_phonon_2007}. It can be realized by concatenating three superlattices with alternating topological invariants, resulting in two interfaces, as illustrated in Fig.~\ref{fig-molecule-design}(a). Each interface supports a localized mode designed at the same frequency. When these modes are coupled, they hybridize, producing a mode splitting that gives rise to symmetric and antisymmetric combinations of the original interface states.

The hybridized modes exhibit overlapping wavefunctions, as evidenced by the displacement fields in Fig.~\ref{fig-molecule-design}(b), which underlies the observed mode splitting. The coupling strength between the interface modes can be tuned by modifying the properties of the central mirror: increasing the number of layers enhances confinement, reducing the mode overlap and the resulting splitting, potentially producing nearly degenerate modes. Quantitatively, the coupling strength is given by $\kappa = \Delta \omega / 2$, where $\Delta \omega$ is the frequency separation between the hybridized modes. The mode splitting can be estimated using a relation analogous to that of coupled double acoustic cavities ~\cite{lanzillotti-kimura_phonon_2007}
\begin{equation}
    \Delta \omega = v_{\textrm{eff}} \frac{\sqrt{1 - \textrm{R}_{\textrm{C}}}}{\pi L_{\mathrm{eff}}},
\end{equation}
where $v_\mathrm{eff}$ is the effective longitudinal sound velocity, $\mathrm{R}_\mathrm{C}$ denotes the reflectivity of the central DBR, and $L_\mathrm{eff}$ is the effective cavity length. The latter exceeds the physical thickness of the central DBR due to evanescent penetration of the acoustic mode into the adjacent reflectors. In the present case, $L_\mathrm{eff}$ is approximated as the central DBR thickness together with an additional contribution corresponding to four unit-cell thicknesses, accounting for penetration on both sides ~\cite{fainsteinRamanScatteringResonant2006}. The analytical expression is plotted in Fig.~\ref{fig-molecule-design}(c), together with transfer-matrix results showing the dependence of the mode splitting on the number of DBR periods.

Time-resolved pump–probe transient reflectivity measurements were performed to characterize the topological phonon molecule. The investigated sample consists of a GaAs/AlAs multilayer structure grown on a GaAs substrate by MBE. The structure contains three superlattices separated by two interfaces. Each of the outer superlattices contains eight periods of a centrosymmetric GaAs/AlAs/GaAs unit cell with layer thicknesses of 6~nm / 4.7~nm / 6~nm, respectively. The central superlattice consists of five periods of the band-inverted unit cell with layer thicknesses of 2~nm / 14.1~nm / 2~nm. The structure is terminated by a 6~nm GaAs cap layer.

\begin{figure*}[!t]
    \includegraphics[width=0.95\textwidth]{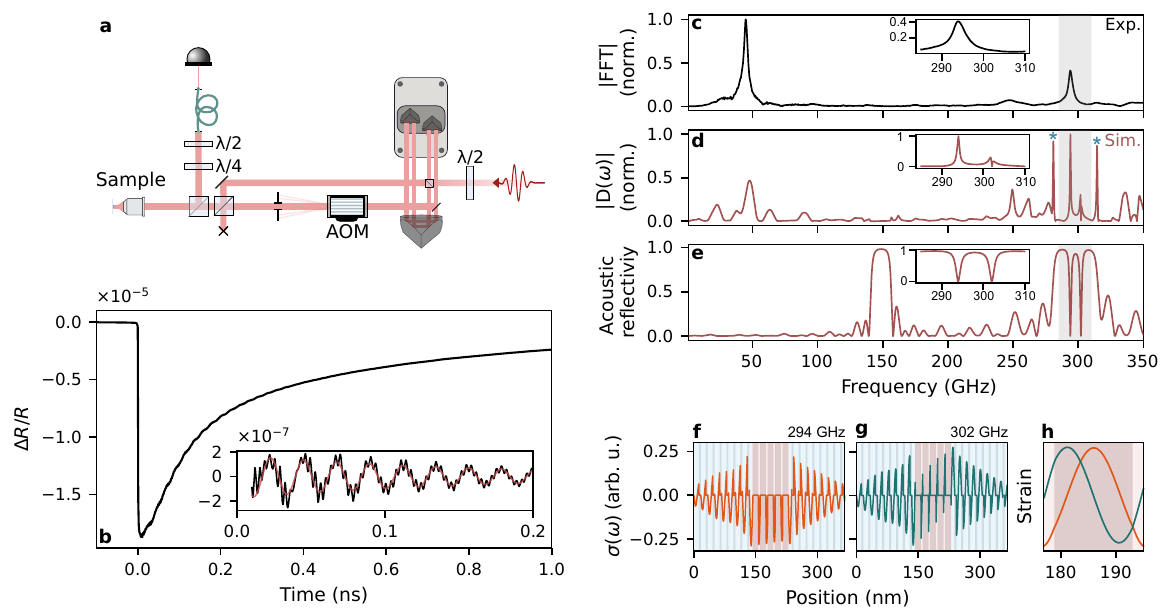}
    \caption{Experimental observation and modeling of the topological phonon molecule.
(a) Schematic of the transient reflectivity pump--probe setup.
(b) Measured transient reflectivity signal; inset: acoustic component after removal of the background. The red line shows a sinusoidal fit to the Brillouin oscillations, with $f_b = 45~\mathrm{GHz}$.
(c) Fourier spectrum of the acoustic component of the signal; inset: zoom highlighting the dominant topological interface mode near $294~\mathrm{GHz}$.
(d) Simulated $D(\omega)$ obtained from the photoelastic model. Star-marked peaks correspond to $q=0$ modes of the individual superlattices and appear attenuated in the experimental spectra. Inset: zoom of the hybridized modes in the same spectral region as the experiment.
(e) Calculated acoustic reflectivity of the multilayer structure. Inset: zoom around the band containing the interface states.
(f,g) Spatial profiles of the integrand of $\sigma(\omega)$ for the two interface modes.
(h) Strain field within the unit cell of the central superlattice, showing symmetric and antisymmetric distributions for the two hybridized modes.}
    \label{fig-molecule-experiment}
\end{figure*}

The experiment was performed using a femtosecond Ti:sapphire laser (Coherent Chameleon Ultra II) with a pulse duration of $\sim$140~fs and a repetition rate of 80~MHz. Pump and probe beams were cross polarized and focused onto the sample at normal incidence using a microscope objective, resulting in a spot diameter of $\sim$10~$\mu$m. The reflected probe beam was detected using a fast photodetector, while residual pump light was suppressed using polarization filtering. Pump-induced reflectivity changes were measured using lock-in detection referenced to the 800~kHz pump modulation frequency provided by an acousto-optic modulator. The pump–probe delay was controlled using a multipath mechanical delay line. The average optical powers were approximately 5~mW for the pump and 2~mW for the probe. The excitation wavelength was 775~nm, above the electron–hole resonance energy of the thinnest quantum well in the structure. A schematic of the experimental setup is shown in Fig.~\ref{fig-molecule-experiment}(a).

The transient reflectivity signal is shown in Fig.~\ref{fig-molecule-experiment}(b). The signal exhibits a sharp change around zero time delay followed by a slow relaxation of the reflectivity, originating from electronic excitation and subsequent thermalization processes induced by the femtosecond pump pulse. To extract the acoustic contribution, this slowly varying background was removed using a Savitzky--Golay smoothing filter. The resulting acoustic contribution to the transient reflectivity signal is shown in the inset of Fig.~\ref{fig-molecule-experiment}(b). It consists of high-frequency oscillations embedded on a slower oscillatory background. The low-frequency oscillation at approximately $45~\mathrm{GHz}$ corresponds to the Brillouin oscillation associated with the propagation of coherent acoustic phonons in the structure. Its frequency is given by
\begin{equation}
    f_B = \frac{2 n_{\mathrm{eff}} v_{\mathrm{eff}}}{\lambda_{\mathrm{probe}}},
\end{equation}

where $n_{\mathrm{eff}}$ and $v_{\mathrm{eff}}$ are the effective refractive index and sound velocity of the structure, respectively.

The frequency spectrum of the high-frequency oscillations, obtained from the fast Fourier transform (FFT) of the acoustic signal, is shown in the inset of Fig.~\ref{fig-molecule-experiment}(c). A dominant spectral peak is observed near $\sim 294~\mathrm{GHz}$, which coincides with a dip in the calculated acoustic reflectivity spectrum shown in Fig.~\ref{fig-molecule-experiment}(e). This peak corresponds to one of the hybridized interface modes of the topological phonon molecule. The measured peak at $\sim 294~\mathrm{GHz}$ is in good agreement with the calculated resonance frequency within a few GHz. The second hybridized mode predicted by the acoustic reflectivity calculation is not observed. As we will demonstrate, this is related to the symmetry of the second peak of the studied modes. We performed additional simulations using a photoelastic model based on the transfer-matrix formalism for one-dimensional multilayer structures. The generation efficiency of coherent acoustic phonons is described by the Brillouin scattering cross section $\sigma(\omega)$, defined as ~\cite{lanzillotti-kimura_theory_2011,PascualWinterSpectralResponsesPhonon2012,ortiz_topological_2021}

\begin{equation}
    \sigma(\omega) = \int |E_{\mathrm{pump}}(z)|^2 \, k(z)\, \frac{\partial u}{\partial z}\, dz ,
\end{equation}

where $\partial u/\partial z$ represents the strain field calculated using the transfer-matrix method under stress-free boundary conditions at the surface of the sample. The term $|E_{\mathrm{pump}}(z)|^2$ denotes the optical intensity distribution of the pump field inside the structure, while $k(z)$ is a material-dependent coefficient describing the efficiency of phonon generation.

The detection efficiency $D(\omega)$ of the probe beam can be approximated as~\cite{ortiz_topological_2021,PascualWinterSpectralResponsesPhonon2012}

\begin{equation}
    D(\omega) = \int \sigma(\omega) E_{\mathrm{probe}}(z)^2 p(z) \frac{\partial u}{\partial z} \, dz ,
\end{equation}

which depends on the strain field, the squared probe electric-field amplitude, and the photoelastic coefficient $p(z)$. Since the transduction efficiency and photoelastic response of GaAs are significantly larger than those of AlAs, a common approximation is to set both $k(z)$ and $p(z)$ equal to unity in GaAs and zero in AlAs~\cite{ortiz_topological_2021}. Although this simplified model neglects detailed material parameters, it captures the essential spectral features of the transient reflectivity signal.

The simulated $D(\omega)$ is shown in Fig.~\ref{fig-molecule-experiment}(d). The calculation reproduces the main experimental observation: a dominant peak associated with the first hybridized interface mode and a strongly suppressed contribution for the second mode. The absence of second peak is consistent with the symmetry-induced suppression predicted by the photoelastic model.

To gain further physical insight, we analyze the spatial dependence of the integrand of $\sigma(\omega)$ for the two interface modes. The corresponding profiles are shown in Figs.~\ref{fig-molecule-experiment}(f,g). In the outer superlattices, the integrand contributions are similar for both modes and largely cancel due to symmetry. The main difference arises in the central superlattice region. 
For the $294~\mathrm{GHz}$ mode, the integrand maintains the same sign throughout the central superlattice,  resulting in efficient excitation of a ‘bright’ mode. In contrast, for the $302~\mathrm{GHz}$ mode, the integrand exhibits both positive and negative contributions, which partially cancel when integrated over the structure, leading to a strongly suppressed (‘dark’) mode. The origin of this difference can be traced to the symmetry of the strain field within the unit cell of the central superlattice. As shown in Fig.~\ref{fig-molecule-experiment}(h), the strain field of the $294~\mathrm{GHz}$ mode is symmetric with respect to the unit cell, whereas the $302~\mathrm{GHz}$ mode exhibits an antisymmetric profile. This antisymmetric character reduces the overlap integral entering $\sigma(\omega)$, making this mode significantly less efficient to excite in the pump--probe experiment.

We note that the experimental FFT peak appears broader than the corresponding simulated spectral response. This difference may arise from several effects that are not included in the present model. In particular, attenuation of high-frequency phonons can contribute to spectral broadening in the experiment. In addition, the finite temporal window of the transient reflectivity measurement imposes a limited frequency resolution, which further broadens the experimental peak. Despite these differences, the simplified simulation reproduces the main spectral features and explains the dominant observation of a strong first hybridized mode and a suppressed second one.
\section{ROBUSTNESS AGAINST THICKNESS-RATIO FLUCTUATIONS}
\begin{figure}[ht]
    \centering
    \includegraphics[width=0.5\columnwidth]{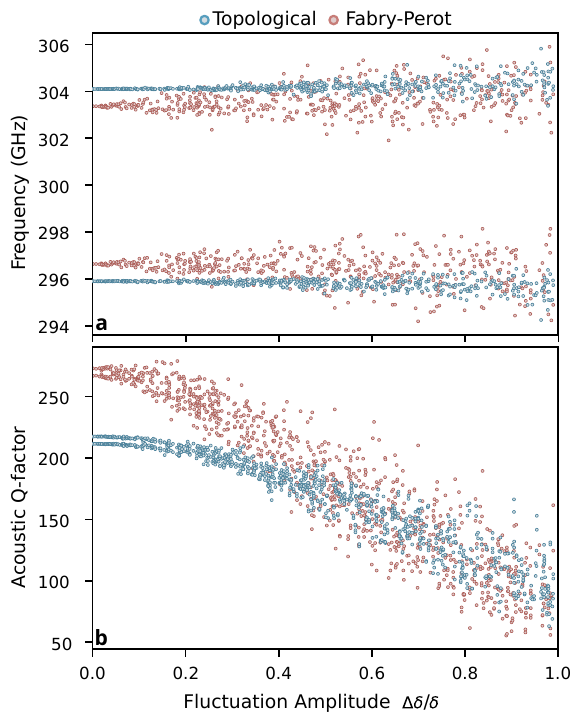}
    \caption{    Comparison of the stability of coupled modes in topological and Fabry-Perot phonon molecules under thickness-ratio fluctuations. The fluctuation amplitude $\Delta\delta$ modifies the GaAs/AlAs thickness ratio and is applied independently to each unit cell using a uniform distribution. Simulated (a) acoustic resonance frequencies and (b) quality factors of the two coupled phonon modes as a function of $\Delta\delta/\delta$. Frequency stability is significantly improved in the topological case, whereas Q-factor degradation remains governed mainly by disorder-induced scattering in both systems.}
    \label{fig:robustness}
\end{figure}

Single interface states produced by band inversion are known to be robust against perturbations that do not close the acoustic band gap or change the Zak-phase associated to the perturbed unit cell~\cite{esmann_topological_2018_1,ortiz_topological_2021}. Here, we examine whether this robustness is preserved when two such interface states are coupled to form a topological phonon molecule. To this end, we compare the response of the coupled topological interface states with that of a conventional Fabry–Pérot phonon molecule under controlled fluctuations of the GaAs/AlAs thickness ratio. The comparison focuses on two experimentally relevant quantities: the resonance frequencies and acoustic quality factors of the two coupled modes.

The centrosymmetric unit cell of the topological structure consists of three layers: $\frac{\lambda_{\text{GaAs}}}{4}(1+\delta)$, $\frac{\lambda_{\text{AlAs}}}{2}(1-\delta)$, and $\frac{\lambda_{\text{GaAs}}}{4}(1+\delta)$. The structure is embedded in a GaAs medium and consists of two 8-period outer DBRs with $\delta=0.5$ and a 5-period center DBR with $\delta=-0.5$. The thickness-ratio fluctuation $\Delta\delta$ is varied from $-0.5$ to $0.5$ and applied independently to each unit cell using a random factor $r_n$ uniformly distributed between 0 and 0.99. Thus, the $n$-th unit cell has layer thicknesses $\frac{\lambda_{\text{GaAs}}}{4}(1+\delta+\Delta\delta r_n)$ , $\frac{\lambda_{\text{AlAs}}}{2}(1-\delta-\Delta\delta r_n)$ , and $\frac{\lambda_{\text{GaAs}}}{4}(1+\delta+\Delta\delta r_n)$. This perturbation changes the relative thickness between the GaAs and AlAs layers within one unit cell while the total acoustic thickness remains constant.
In the Fabry-Perot case, the non-centrosymmetric unit cell consists of two layers: $\frac{\lambda_{\text{GaAs}}}{2}(1+\delta)$ and $\frac{\lambda_{\text{AlAs}}}{2}(1-\delta)$. The full structure is composed of three DBRs with 8, 5, and 8 periods ($\delta=0.5$), respectively, separated by two $\lambda$-thick spacer layers. The spacer thicknesses are kept fixed, while the relative layer thicknesses within the DBRs are perturbed in the same way as in the topological structure.

The results are summarized in Fig.~\ref{fig:robustness}. For the topological molecule, the two resonance frequencies remain clustered around their unperturbed values over the full range of thickness-ratio fluctuations, whereas the Fabry–Pérot molecule exhibits a broader frequency dispersion. This behavior reflects the fact that the topological modes remain pinned to the interface-state frequency as long as the perturbation does not close the relevant band gap or remove the band inversion. On the contrary, the perturbations in the central superlattice affects its reflectivity, and thus the coupling factor between the two modes. This results in a perturbation dependent coupling.  It must be noted however, that while not strictly robust against thickness ratio fluctuations, the modes are more stable than in the case of the Fabry–Pérot molecule.  The quality factors decrease with increasing disorder amplitude in both systems, as expected from enhanced scattering and imperfect confinement. However, the topological molecule preserves a more stable spectral response, indicating that the coupled modes retain the localization associated with the underlying Zak-phase contrast.

These simulations therefore support the interpretation of the hybridized modes as coupled topological interface states rather than conventional coupled cavity modes. The topology does not prevent all disorder-induced loss, but it stabilizes the resonance frequencies and preserves interface localization over a broad range of thickness-ratio perturbations.

%\section{Engineering multiple interface %states} \label{topological chain}

\section{From molecules to chains: multiple coupled interface states} \label{topological chain}
The two-interface system discussed above naturally generalizes to a chain of N coupled topological interface states ~\cite{lanzillotti-kimura_phonon_2007}, forming a one-dimensional lattice of interacting topologically localized modes. In analogy with coupled cavity arrays, a sequence of alternating topological superlattices generates a chain of localized interface states whose mutual coupling leads to the formation of collective modes. In this picture, each interface state plays the role of an effective phononic ``site'', while the superlattices separating neighboring interfaces control the coupling strength between them.

\begin{figure}[ht]
    \centering
    \includegraphics[width=0.6\columnwidth]{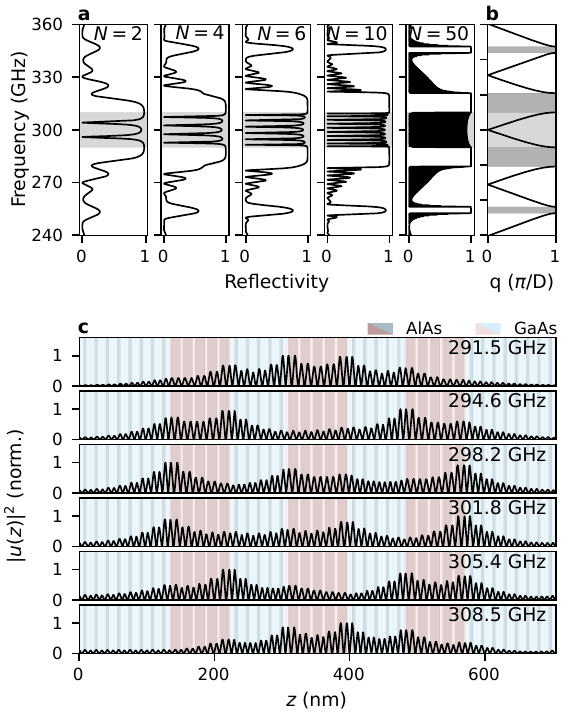}
    \caption{Phononic chains formed by coupled topological interface states.
(a) Acoustic reflectivity  for structures (top-left) with different numbers of interfaces $N$, illustrating the evolution from isolated interface modes to a miniband.
(b) Phononic dispersion (top-right) for the infinite-chain limit, showing a miniband (light shaded) arising from coupled interface states; bandgaps are indicated by dark-shaded regions.
(c) Spatial profiles of the squared displacement $|u(z)|^2$ for the case $N=6$, showing the six hybridized interface modes localized at the interfaces.}
    \label{fig-chain-design}
\end{figure}

This evolution is illustrated in Fig.~\ref{fig-chain-design}(a), which shows the calculated acoustic reflectivity for structures containing different numbers of interfaces $N$. Each interface is formed by joining two band-inverted superlattices; the structure comprises two outer DBRs with 8 periods each and $N-1$ inner DBRs of 5 periods, based on the same unit cells as described previously. For $N=2$, the system corresponds to the phonon molecule discussed previously, characterized by two hybridized modes resulting from the coupling of the two interface states. Increasing the number of interfaces leads to the appearance of additional resonances inside the band gap corresponding to the collective modes of the coupled system. As $N$ increases further, these discrete resonances progressively form a narrow miniband centered around the original interface-state frequency. In the infinite-chain limit, this miniband converges to the allowed band given by the phononic dispersion relation [Fig.~\ref{fig-chain-design}(b)], arising from the periodic coupling of interface states. The dispersion is calculated for a periodic structure composed of band-inverted DBRs with a periodicity of five, by applying Bloch boundary conditions as described in the Appendix.

The spatial profiles of the displacement field for the case $N=6$ are shown in Fig.~\ref{fig-chain-design}(c). The six eigenmodes correspond to different collective combinations of the localized interface states. Although the modes extend over the entire structure, their amplitude remains strongly concentrated near the interfaces, reflecting the localized nature of the underlying topological states. The relative phase between neighboring interfaces determines the symmetry of the different modes, in analogy with the formation of molecular orbitals in coupled systems.

The bandwidth of the resulting miniband is controlled by the effective coupling between neighboring interface states, which depends on the reflectivity and thickness of the separating superlattices. Increasing the separation between interfaces reduces the overlap of the localized modes and therefore decreases the coupling strength, leading to a narrower miniband. In the limit of large separations, the modes recover the behavior of isolated interface states.

This system therefore provides a topological analogue of phononic chains previously realized theoretically using coupled cavity spacers~\cite{lanzillotti-kimura_phonon_2007}. In contrast to conventional Fabry--P\'erot cavities, the localized modes in the present structure originate from topological band inversion and remain pinned within the band gap of the surrounding superlattices. Consequently, the collective modes formed in the chain benefit from the spatial localization and robustness of the individual interface states while allowing controllable coupling between them.

\begin{figure}[ht]
    \centering
    \includegraphics[width=0.6\columnwidth]{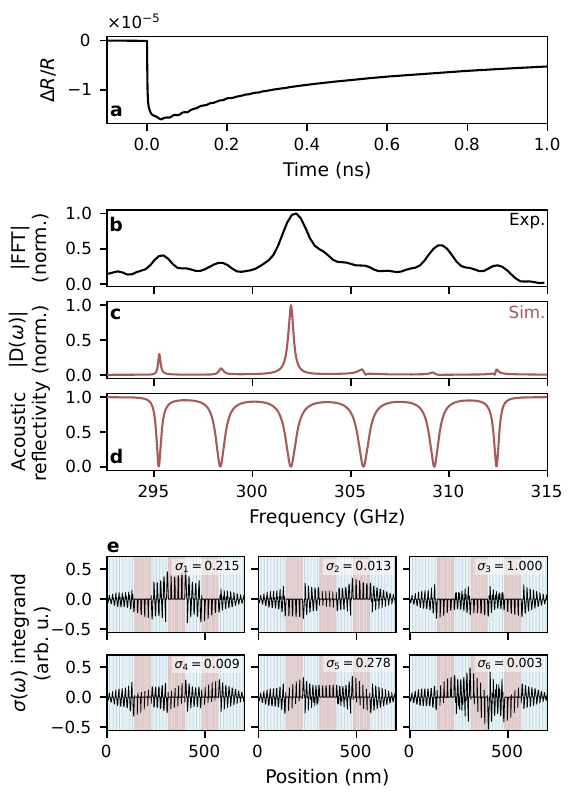}
    \caption{Experimental observation and modeling of the topological phonon chain.
(a) Transient reflectivity signal measured for the sample containing six interfaces.
(b) Fourier spectrum of the acoustic component after removal of the electronic background and Brillouin oscillation.
(c) Simulated $D(\omega)$ obtained from the photoelastic model.
(d) Calculated acoustic reflectivity of the multilayer structure in the spectral region of the interface modes.
(e) Spatial dependence of the integrand of $\sigma(\omega)$ for the different interface modes.}
    \label{fig-chain-experiment}
\end{figure}

To experimentally investigate the topological phonon chain, transient reflectivity pump--probe measurements were performed on a multilayer structure containing six interfaces. The outer superlattices consist of eight periods of a centrosymmetric GaAs/AlAs/GaAs unit cell with layer thicknesses of 6~nm / 4.7~nm / 6~nm. The internal section of the structure is composed of alternating superlattices formed by five periods of the unit cell (2~nm / 14.1~nm / 2~nm) and five periods of the unit cell (6~nm / 4.7~nm / 6~nm), repeated three and two times, respectively. The structure is capped with a 6~nm GaAs layer. The experimental conditions are identical to those used for the topological phonon molecule sample described in the previous section. 
The measured transient reflectivity signal for the sample is shown in Fig.~\ref{fig-chain-experiment}(a). After removing the slowly varying electronic and thermal background together with the low-frequency Brillouin oscillation, the acoustic component of the signal was analyzed in the frequency domain. The corresponding Fourier spectrum is shown in Fig.~\ref{fig-chain-experiment}(b), where a dominant peak is observed near $302~\mathrm{GHz}$ together with weaker side peaks. To interpret these observations, we calculated the $D(\omega)$ of the structure using the photoelastic model described previously. The simulated response in Fig.~\ref{fig-chain-experiment}(c) shows a dominant peak near $302~\mathrm{GHz}$ corresponding to the third dip in the calculated acoustic reflectivity spectrum. The simulation also predicts weaker contributions from the remaining interface modes. The dominant experimental peak observed in Fig.~\ref{fig-chain-experiment}(b), together with the weaker side peaks, aligns well with the reflectivity features in Fig.~\ref{fig-chain-experiment}(d), confirming that the detected resonances originate from the coupled interface states of the topological chain.
Although the relative amplitudes of the peaks in the experiment do not exactly match the simulated spectrum, this discrepancy can be attributed to the approximations used in the calculations. In these simulations, $\sigma(\omega)$ and $D(\omega)$ were evaluated using the same approach as in the two-interface sample. A more quantitative description would require explicit modeling of the photoelastic coefficients and phonon generation efficiency in each individual layer~\cite{mlayahRamanBrillouinLightScattering2007,groenenInelasticLightScattering2008,jusserandSelectiveResonantInteraction2013,jusserandPolaritonResonancesUltrastrong2015}. Such a detailed treatment was not pursued here, since the simplified model already reproduces the main experimental observations.

Insight into the dominant excitation of the third interface mode is obtained by examining the spatial profile of the integrand of the Brillouin cross section shown in Fig.~\ref{fig-chain-experiment}(e). The different panels correspond to the spatial distribution of the integrand of $\sigma(\omega)$ for the different interface modes. For the third mode (Fig.~\ref{fig-chain-experiment}(e), top-right panel), the contributions from alternating inner superlattices of the structure have the same sign and therefore add constructively in the overlap integral, leading to efficient phonon generation. This behavior arises from the symmetric nature of the strain profile of this mode within the corresponding superlattice unit cells, similar to the mechanism discussed for the phonon molecule case. In contrast, the other modes exhibit combinations of symmetric and antisymmetric strain profiles across the different superlattices, or symmetric strain profiles with alternating signs. As a result, the integrand contributions partially cancel in the overlap integral, leading to a reduced generation efficiency for these modes. The relative values of $\sigma(\omega)$ for the different modes are indicated in the figure. In the infinite-chain limit, the phononic dispersion [Fig.~\ref{fig-chain-design}(b)] shows that the miniband is centered at the interface-state frequency, with the q=0 mode located at the design frequency. Modes near q=0 are preferentially generated and detected in pump–probe experiments \cite{PascualWinterSpectralResponsesPhonon2012}.

\section{Conclusion and perspectives} \label{conclusion}
We have investigated the formation and coupling of topological phononic interface states in GaAs/AlAs superlattices engineered through band inversion of centrosymmetric unit cells. Two coupled interfaces give rise to hybridized modes forming a phononic molecule, characterized by symmetric and antisymmetric combinations of the localized interface states within the acoustic band gap. These hybridized states were experimentally observed using transient reflectivity pump--probe measurements, with resonance frequencies consistent with the calculated acoustic reflectivity spectrum and $D(\omega)$. We further extended this concept to structures containing multiple interfaces, forming chains of coupled topological interface states. As the number of interfaces increases, the localized modes hybridize into a set of collective resonances that evolve toward a miniband structure. Experimental measurements on a chain sample containing six interfaces reveal a dominant resonance consistent with the calculated detection response and the corresponding acoustic reflectivity features of the structure.

Analysis of the spatial profile of the Brillouin generation integrand shows that the excitation efficiency of the different modes is governed by the symmetry of the strain field and the resulting constructive or destructive contributions in the overlap integral. In particular, the dominant excitation of one interface mode arises from constructive contributions across the multilayer structure, while other modes are suppressed due to partial cancellation of the generation integrand. These results demonstrate the controlled realization of phononic molecules and chains based on topological interface states in semiconductor superlattices. Altogether, these findings establish topological interface states in semiconductor superlattices as a versatile platform for engineering coupled phononic systems. The realization of controllable phononic molecules and chains opens new possibilities for manipulating coherent acoustic phonons and designing complex nanophononic structures.

Combined with the atomic-scale thickness control provided by MBE, these architectures can set a basis for constructing robust cavity modes using quasiperiodic arrays and for engineering complex phononic band structures spanning the gigahertz to terahertz frequency range. Owing to the “magic” coincidence that GaAs/AlAs stacks with appropriately tuned layer thicknesses act as efficient Bragg mirrors for both near-infrared light and gigahertz acoustic phonons, the design principles developed here are directly translatable to optophononic structures, where coupled topological phonon molecules can be co-localized with optical cavity or interface modes. These results establish coupled topological interface states as a controllable route to engineer phononic molecules and minibands in semiconductor superlattices, with direct prospects for GHz optophononic devices.

\section {acknowledgement}
The authors acknowledge funding from European Research Council Consolidator Grant No.101045089 (T-Recs). The authors acknowledge the French RENATECH network which partly funds the MBE growth.

\section {APPENDIX: CALCULATION OF THE ZAK PHASE} \label{appendix_zak}

\paragraph{Eigenvalue problem of 1D superlattice} The propagation of longitudinal acoustic waves in a periodic one-dimensional structure is described by the elastic wave equation. The acoustic field vector ($U(z), S(z)$) at the left and right boundaries of layer $i$ is related by the transfer matrix $M_i$:~\cite{tamuraAcousticphononTransmissionQuasiperiodic1987,lanzillotti-kimura_phonon_2007}
\setcounter{equation}{0}
\renewcommand{\theequation}{A\arabic{equation}}

\begin{equation}
M_i =
\begin{pmatrix}
\cos(k_i d_i) & \dfrac{\sin(k_i d_i)}{\omega Z_i} \\
-\omega Z_i \sin(k_i d_i) & \cos(k_i d_i)
\end{pmatrix}
\label{eq:A1}
\end{equation}
where $k_i$ is the wave vector, $d_i$ is the layer thickness, $\omega$ is the angular frequency, and $Z_i$ is the acoustic impedance. For a unit cell composed of multiple layers with total thickness $d$, The overall transfer matrix $M$ is obtained by multiplying the individual layer matrices. The acoustic field across the unit cell can then be written as
\begin{equation}
\begin{pmatrix}
U(d) \\
S(d)
\end{pmatrix}
=
M
\begin{pmatrix}
U(0) \\
S(0)
\end{pmatrix}.
\label{eq:A2}
\end{equation}
For an infinite periodic structure, the dispersion relation is obtained by imposing the Bloch condition on the acoustic field,

\begin{equation}
\begin{pmatrix}
U(d) \\
S(d)
\end{pmatrix}
=
e^{iqd}
\begin{pmatrix}
U(0) \\
S(0)
\end{pmatrix},
\label{eq:A3}
\end{equation}

where $q$ is the Bloch wave vector. Combining Eqs.~(\ref{eq:A2}) and Eqs.~(\ref{eq:A3}), one obtains an eigenvalue problem for $M$, whose eigenvalues are given by $\lambda_{\pm} = e^{\pm i q d}$. Solving this eigenvalue problem gives the dispersion relation and Bloch eigenvectors $(U(0), S(0))$, which define the displacement and stress at the unit-cell boundary. The Bloch eigenfunctions are obtained by propagating the state vector through successive layers using the transfer matrix.

\paragraph{Calculating the Zak phase} The Zak phase is the geometric phase that characterizes the topological properties of bands in one-dimensional periodic systems. In inversion-symmetric systems, the Zak phase is quantized to 0 or $\pi$ and is determined by the symmetry of the Bloch eigenfunctions at the band edges. When the eigenfunctions at the center $(q=0)$ and boundary $(q=\pi/d)$ of the Brillouin zone have the same symmetry, the Zak phase of the $n$-th band is $\theta_n^{\mathrm{Zak}} = 0$. In contrast, when these modes have opposite symmetries, the Zak phase is $\theta_n^{\mathrm{Zak}} = \pi$.  For the \(n\)-th band, the Zak phase is defined as ~\cite{xiao_geometric_2015}
\begin{equation}
\begin{split}
\theta_n^{\text{Zak}}
=
\int_{-\pi/d}^{\pi/d}
\left[
i
\int_{\text{unit\ cell}}
\frac{1}{2\rho(z)v^2(z)}\,
u_{n,q}^*(z)\,
\partial_q u_{n,q}(z)\,
\mathrm dz
\right]
\mathrm dq
\end{split}
\end{equation}

Here, $u_{n,q}(z)$ is the normalized Bloch eigenfunction, and $\rho(z)$ and $v(z)$ are the position-dependent mass density and wave velocity within the unit cell, respectively.
\setcounter{figure}{0}
\renewcommand{\thefigure}{A\arabic{figure}}
\begin{figure}
    \centering
    \includegraphics[width=0.5\linewidth]{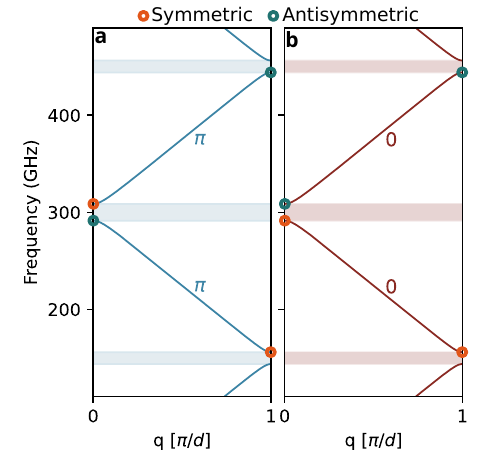}
    \caption{Acoustic dispersion relations and corresponding Zak phases for two centrosymmetric superlattices: (a) $(3\lambda_{\mathrm{GaAs}}/8,\ \lambda_{\mathrm{AlAs}}/4,\ 3\lambda_{\mathrm{GaAs}}/8)$ and (b) $(\lambda_{\mathrm{GaAs}}/8,\ 3\lambda_{\mathrm{AlAs}}/4,\ \lambda_{\mathrm{GaAs}}/8)$. In (a), the eigenfunctions at $q=0$ and $q=\pi/d$ have opposite symmetries in both branches around 300~GHz, giving $\theta^{\mathrm{Zak}}=\pi$, whereas in (b) the symmetry is preserved, yielding $\theta^{\mathrm{Zak}}=0$. The resulting Zak phase  contrast leads to a localized interface state when two structure joined. Markers indicate band-edge symmetries as in Fig.~\ref{fig-interface}.}
    \label{fig:zak_phase}
\end{figure}

For numerical calculation, the folded Brillouin zone is sampled at discrete wave vectors \(q_i\), and the Zak phase is calculated using the discrete form~\cite{xiao_geometric_2015}
\begin{equation}
\begin{split}
\theta_n^{\text{Zak}}
=
-\text{Im}
\sum_{i=1}^{N}
\ln
\left[
\int_{\text{unit cell}}
\frac{1}{2\rho(z)v^2(z)}
u_{n,q_i}^*(z)\,
u_{n,q_{i+1}}(z)\,
\mathrm dz
\right]
\end{split}
\end{equation}

In this expression, \(q_i\) and \(q_{i+1}\) are two neighboring sampling points in the Brillouin zone, and the integral gives the overlap between adjacent Bloch eigenmodes within the same band. Figure~A1 shows the acoustic dispersion relations and corresponding Zak phases for two centrosymmetric superlattices presented in the main text. The Zak phase of the bands changes from $\pi$ to $0$, reflecting a band inversion and the associated topological transition.

\bibliographystyle{plain}
\bibliography{references}

@article{lanzillotti-kimura_bloch_2010,
	title = {Bloch {Oscillations} of {THz} {Acoustic} {Phonons} in {Coupled} {Nanocavity} {Structures}},
	volume = {104},
	url = {https://link.aps.org/doi/10.1103/PhysRevLett.104.197402},
	doi = {10.1103/PhysRevLett.104.197402},

	number = {19},
	urldate = {2023-01-23},
	journal = {Physical Review Letters},
	author = {Lanzillotti-Kimura, N. D. and Fainstein, A. and Perrin, B. and Jusserand, B. and Mauguin, O. and Largeau, L. and Lemaître, A.},
	month = may,
	year = {2010},
	pages = {197402},
}

@article{esmann_topological_2018_2,
	title = {A {Topological} {View} on {Optical} and {Phononic} {Fabry}–{Perot} {Microcavities} through the {Su}–{Schrieffer}–{Heeger} {Model}},
	volume = {8},
	copyright = {http://creativecommons.org/licenses/by/3.0/},
	issn = {2076-3417},
	url = {https://www.mdpi.com/2076-3417/8/4/527},
	doi = {10.3390/app8040527},
	
	number = {4},
	urldate = {2023-01-17},
	journal = {Applied Sciences},
	author = {Esmann, Martin and Lanzillotti-Kimura, Norberto Daniel},
	month = apr,
	year = {2018},
	pages = {527},
}

@article{zhang_topological_2018,
	title = {Topological sound},
	volume = {1},
	copyright = {2018 The Author(s)},
	issn = {2399-3650},
	url = {https://www.nature.com/articles/s42005-018-0094-4},
	doi = {10.1038/s42005-018-0094-4},

	
	number = {1},
	urldate = {2022-12-23},
	journal = {Communications Physics},
	author = {Zhang, Xiujuan and Xiao, Meng and Cheng, Ying and Lu, Ming-Hui and Christensen, Johan},
	month = dec,
	year = {2018},
	pages = {1--13},
}

@article{ortiz_phonon_2019,
	title = {Phonon engineering with superlattices: {Generalized} nanomechanical potentials},
	volume = {100},
	shorttitle = {Phonon engineering with superlattices},
	url = {https://link.aps.org/doi/10.1103/PhysRevB.100.085430},
	doi = {10.1103/PhysRevB.100.085430},
	number = {8},
	urldate = {2022-11-18},
	journal = {Physical Review B},
	author = {Ortiz, O. and Esmann, M. and Lanzillotti-Kimura, N. D.},
	month = aug,
	year = {2019},
	pages = {085430},
}

@article{kettler_inducing_2021,
	title = {Inducing micromechanical motion by optical excitation of a single quantum dot},
	volume = {16},
	copyright = {2020 The Author(s), under exclusive licence to Springer Nature Limited},
	issn = {1748-3395},
	url = {https://www.nature.com/articles/s41565-020-00814-y},
	doi = {10.1038/s41565-020-00814-y},
	number = {3},
	urldate = {2022-11-15},
	journal = {Nature Nanotechnology},
	author = {Kettler, Jan and Vaish, Nitika and de Lépinay, Laure Mercier and Besga, Benjamin and de Assis, Pierre-Louis and Bourgeois, Olivier and Auffèves, Alexia and Richard, Maxime and Claudon, Julien and Gérard, Jean-Michel and Pigeau, Benjamin and Arcizet, Olivier and Verlot, Pierre and Poizat, Jean-Philippe},
	month = mar,
	year = {2021},
	pages = {283--287},
}

@article{stiller_coherently_2020,
author = {Birgit Stiller and Moritz Merklein and Christian Wolff and Khu Vu and Pan Ma and Stephen J. Madden and Benjamin J. Eggleton},
journal = {Optica},
number = {5},
pages = {492--497},
publisher = {Optica Publishing Group},
title = {Coherently refreshing hypersonic phonons for light storage},
volume = {7},
month = {May},
year = {2020},
url = {https://opg.optica.org/optica/abstract.cfm?URI=optica-7-5-492},
doi = {10.1364/OPTICA.386535},
}

@article{lanzillotti-kimura_phonon_2007,
	title = {Phonon engineering with acoustic nanocavities: {Theoretical} considerations on phonon molecules, band structures, and acoustic {Bloch} oscillations},
	volume = {75},
	shorttitle = {Phonon engineering with acoustic nanocavities},
	url = {https://link.aps.org/doi/10.1103/PhysRevB.75.024301},
	doi = {10.1103/PhysRevB.75.024301},
	number = {2},
	urldate = {2022-11-09},
	journal = {Physical Review B},
	author = {Lanzillotti-Kimura, N. D. and Fainstein, A. and Balseiro, C. A. and Jusserand, B.},
	month = jan,
	year = {2007},
	pages = {024301},
}

@article{esmann_topological_2018,
	title = {Topological acoustics in coupled nanocavity arrays},
	volume = {98},
	url = {https://link.aps.org/doi/10.1103/PhysRevB.98.161109},
	doi = {10.1103/PhysRevB.98.161109},
	number = {16},
	urldate = {2022-10-19},
	journal = {Physical Review B},
	author = {Esmann, M. and Lamberti, F. R. and Lemaître, A. and Lanzillotti-Kimura, N. D.},
	month = oct,
	year = {2018},
	pages = {161109(R)},
}

@article{hauff_chiral_2022,
	title = {Chiral quantum optics in broken-symmetry and topological photonic crystal waveguides},
	volume = {4},
	url = {https://link.aps.org/doi/10.1103/PhysRevResearch.4.023082},
	doi = {10.1103/PhysRevResearch.4.023082},
	number = {2},
	urldate = {2022-10-19},
	journal = {Physical Review Research},
	author = {Hauff, Nils Valentin and Le Jeannic, Hanna and Lodahl, Peter and Hughes, Stephen and Rotenberg, Nir},
	month = apr,
	year = {2022},
	pages = {023082},
}

@article{lu_topological_2014,
	title = {Topological photonics},
	volume = {8},
	
	issn = {1749-4893},
	url = {https://www.nature.com/articles/nphoton.2014.248},
	doi = {10.1038/nphoton.2014.248},
	number = {11},
	urldate = {2022-10-19},
	journal = {Nature Photonics},
	author = {Lu, Ling and Joannopoulos, John D. and Soljačić, Marin},
	month = nov,
	year = {2014},
	pages = {821--829},
}

@article{hafezi_imaging_2013,
	title = {Imaging topological edge states in silicon photonics},
	volume = {7},
	copyright = {2013 Nature Publishing Group},
	issn = {1749-4893},
	url = {https://www.nature.com/articles/nphoton.2013.274},
	doi = {10.1038/nphoton.2013.274},
	number = {12},
	urldate = {2022-10-19},
	journal = {Nature Photonics},
	author = {Hafezi, M. and Mittal, S. and Fan, J. and Migdall, A. and Taylor, J. M.},
	month = dec,
	year = {2013},
	pages = {1001--1005},
}

@article{st-jean_measuring_2021,
	title = {Measuring {Topological} {Invariants} in a {Polaritonic} {Analog} of {Graphene}},
	volume = {126},
	url = {https://link.aps.org/doi/10.1103/PhysRevLett.126.127403},
	doi = {10.1103/PhysRevLett.126.127403},
	number = {12},
	urldate = {2022-10-19},
	journal = {Physical Review Letters},
	author = {St-Jean, P. and Dauphin, A. and Massignan, P. and Real, B. and Jamadi, O. and Milicevic, M. and Lema{\^i}tre, A. and Harouri, A. and Le Gratiet, L. and Sagnes, I. and Ravets, S. and Bloch, J. and Amo, A.},
	month = mar,
	year = {2021},
	pages = {127403},
}

@article{pernet_gap_2022,
	title = {Gap solitons in a one-dimensional driven-dissipative topological lattice},
	volume = {18},
	copyright = {2022 The Author(s), under exclusive licence to Springer Nature Limited},
	issn = {1745-2481},
	url = {https://www.nature.com/articles/s41567-022-01599-8},
	doi = {10.1038/s41567-022-01599-8},
	number = {6},
	urldate = {2022-10-19},
	journal = {Nature Physics},
	author = {Pernet, Nicolas and St-Jean, Philippe and Solnyshkov, Dmitry D. and Malpuech, Guillaume and Carlon Zambon, Nicola and Fontaine, Quentin and Real, Bastian and Jamadi, Omar and Lemaître, Aristide and Morassi, Martina and Le Gratiet, Luc and Baptiste, Téo and Harouri, Abdelmounaim and Sagnes, Isabelle and Amo, Alberto and Ravets, Sylvain and Bloch, Jacqueline},
	month = jun,
	year = {2022},
	pages = {678--684},
}

@article{klembt_exciton-polariton_2018,
	title = {Exciton-polariton topological insulator},
	volume = {562},
	copyright = {2018 Springer Nature Limited},
	issn = {1476-4687},
	url = {https://www.nature.com/articles/s41586-018-0601-5},
	doi = {10.1038/s41586-018-0601-5},
	number = {7728},
	urldate = {2022-10-19},
	journal = {Nature},
	author = {Klembt, S. and Harder, T. H. and Egorov, O. A. and Winkler, K. and Ge, R. and Bandres, M. A. and Emmerling, M. and Worschech, L. and Liew, T. C. H. and Segev, M. and Schneider, C. and Höfling, S.},
	month = oct,
	year = {2018},
	pages = {552--556},
}

@article{ozawa_topological_2019,
	title = {Topological photonics},
	volume = {91},
	url = {https://link.aps.org/doi/10.1103/RevModPhys.91.015006},
	doi = {10.1103/RevModPhys.91.015006},
	number = {1},
	urldate = {2022-10-19},
	journal = {Reviews of Modern Physics},
	author = {Ozawa, Tomoki and Price, Hannah M. and Amo, Alberto and Goldman, Nathan and Hafezi, Mohammad and Lu, Ling and Rechtsman, Mikael C. and Schuster, David and Simon, Jonathan and Zilberberg, Oded and Carusotto, Iacopo},
	month = mar,
	year = {2019},
	pages = {015006},
}

@article{ma_topological_2019,
	title = {Topological phases in acoustic and mechanical systems},
	volume = {1},
	issn = {2522-5820},
	url = {https://www.nature.com/articles/s42254-019-0030-x},
	doi = {10.1038/s42254-019-0030-x},
	number = {4},
	urldate = {2022-10-17},
	journal = {Nature Reviews Physics},
	author = {Ma, Guancong and Xiao, Meng and Chan, C. T.},
	month = apr,
	year = {2019},
	pages = {281--294},
}

@book{asboth_short_2016,
	series = {Lecture {Notes} in {Physics}},
	title = {A {Short} {Course} on {Topological} {Insulators}},
	url = {https://link.springer.com/book/10.1007/978-3-319-25607-8},
    publisher = {Springer},
	number = {919},
	urldate = {2022-10-17},
	author = {Asbóth, J.K. and Oroszlány, L. and Pályi, A.},
	year = {2016},
}

@article{su_solitons_1979,
	title = {Solitons in {Polyacetylene}},
	volume = {42},
	url = {https://link.aps.org/doi/10.1103/PhysRevLett.42.1698},
	doi = {10.1103/PhysRevLett.42.1698},
	number = {25},
	urldate = {2022-10-08},
	journal = {Physical Review Letters},
	author = {Su, W. P. and Schrieffer, J. R. and Heeger, A. J.},
	month = jun,
	year = {1979},
	pages = {1698--1701},
}

@article{lanzillotti-kimura_theory_2011,
	title = {Theory of coherent generation and detection of {THz} acoustic phonons using optical microcavities},
	volume = {84},
	issn = {1098-0121, 1550-235X},
	url = {https://link.aps.org/doi/10.1103/PhysRevB.84.064307},
	doi = {10.1103/PhysRevB.84.064307},
	
	number = {6},
	urldate = {2022-07-21},
	journal = {Physical Review B},
	author = {Lanzillotti-Kimura, N. D. and Fainstein, A. and Perrin, B. and Jusserand, B.},
	month = aug,
	year = {2011},
	pages = {064307},
}

@article{doster_observing_2022,
	title = {Observing polarization patterns in the collective motion of nanomechanical arrays},
	volume = {13},
	issn = {2041-1723},
	url = {https://www.nature.com/articles/s41467-022-30024-0},
	doi = {10.1038/s41467-022-30024-0},
	number = {1},
	urldate = {2022-05-31},
	journal = {Nature Communications},
	author = {Doster, Juliane and Shah, Tirth and Fösel, Thomas and Paulitschke, Philipp and Marquardt, Florian and Weig, Eva M.},
	month = dec,
	year = {2022},
	pages = {2478},
}

@article{chu_creation_2018,
	title = {Creation and control of multi-phonon {Fock} states in a bulk acoustic-wave resonator},
	volume = {563},
	issn = {0028-0836, 1476-4687},
	url = {http://www.nature.com/articles/s41586-018-0717-7},
	doi = {10.1038/s41586-018-0717-7},
	
	number = {7733},
	urldate = {2022-05-05},
	journal = {Nature},
	author = {Chu, Yiwen and Kharel, Prashanta and Yoon, Taekwan and Frunzio, Luigi and Rakich, Peter T. and Schoelkopf, Robert J.},
	month = nov,
	year = {2018},
	pages = {666--670},
}

@article{ortiz_topological_2021,
	title = {Topological optical and phononic interface mode by simultaneous band inversion},
	volume = {8},
	issn = {2334-2536},
	url = {https://opg.optica.org/abstract.cfm?URI=optica-8-5-598},
	doi = {10.1364/OPTICA.411945},
	
	number = {5},
	urldate = {2022-02-15},
	journal = {Optica},
	author = {Ortiz, O. and Priya, P. and Rodriguez, A. and Lema{\^i}tre, A. and Esmann, M. and Lanzillotti-Kimura, N. D.},
	month = may,
	year = {2021},
	pages = {598},
}

@article{ortiz_fiber-integrated_2020,
	title = {Fiber-integrated microcavities for efficient generation of coherent acoustic phonons},
	volume = {117},
	issn = {0003-6951},
	url = {https://aip.scitation.org/doi/abs/10.1063/5.0026959},
	doi = {10.1063/5.0026959},
	number = {18},
	urldate = {2021-04-12},
	journal = {Applied Physics Letters},
	author = {Ortiz, O. and Pastier, F. and Rodriguez, A. and {Priya} and Lema{\^i}tre, A. and Gomez-Carbonell, C. and Sagnes, I. and Harouri, A. and Senellart, P. and Giesz, V. and Esmann, M. and Lanzillotti-Kimura, N. D.},
	month = nov,
	year = {2020},

	pages = {183102},
}

@article{chafatinos_polariton-driven_2020,
	title = {Polariton-driven phonon laser},
	volume = {11},
	issn = {2041-1723},
	url = {http://www.nature.com/articles/s41467-020-18358-z},
	doi = {10.1038/s41467-020-18358-z},
	number = {1},
	urldate = {2020-10-08},
	journal = {Nature Communications},
	author = {Chafatinos, D. L. and Kuznetsov, A. S. and Anguiano, S. and Bruchhausen, A. E. and Reynoso, A. A. and Biermann, K. and Santos, P. V. and Fainstein, A.},
	month = dec,
	year = {2020},
	pages = {4552},
}

@article{arregui_coherent_2019,
	title = {Coherent generation and detection of acoustic phonons in topological nanocavities},
	volume = {4},
	issn = {2378-0967},
	url = {http://aip.scitation.org/doi/10.1063/1.5082728},
	doi = {10.1063/1.5082728},
	number = {3},
	urldate = {2020-10-01},
	journal = {APL Photonics},
	author = {Arregui, G. and Ortíz, O. and Esmann, M. and Sotomayor-Torres, C. M. and Gomez-Carbonell, C. and Mauguin, O. and Perrin, B. and Lemaître, A. and García, P. D. and Lanzillotti-Kimura, N. D.},
	month = mar,
	year = {2019},
	pages = {030805},
}

@article{esmann_brillouin_2019,
	title = {Brillouin scattering in hybrid optophononic {Bragg} micropillar resonators at 300 {GHz}},
	volume = {6},
	issn = {2334-2536},
	url = {https://www.osapublishing.org/abstract.cfm?URI=optica-6-7-854},
	doi = {10.1364/OPTICA.6.000854},
	
	number = {7},
	urldate = {2020-10-01},
	journal = {Optica},
	author = {Esmann, M. and Lamberti, F. R. and Harouri, A. and Lanco, L. and Sagnes, I. and Favero, I. and Aubin, G. and Gomez-Carbonell, C. and Lemaître, A. and Krebs, O. and Senellart, P. and Lanzillotti-Kimura, N. D.},
	month = jul,
	year = {2019},
	pages = {854},
}

@article{xiao_geometric_2015,
	title = {Geometric phase and band inversion in periodic acoustic systems},
	volume = {11},
	issn = {1745-2473, 1745-2481},
	url = {http://www.nature.com/articles/nphys3228},
	doi = {10.1038/nphys3228},
	
	number = {3},
	urldate = {2020-09-29},
	journal = {Nature Physics},
	author = {Xiao, Meng and Ma, Guancong and Yang, Zhiyu and Sheng, Ping and Zhang, Z. Q. and Chan, C. T.},
	month = mar,
	year = {2015},
	pages = {240--244},
}

@article{xiao_surface_2014,
	title = {Surface {Impedance} and {Bulk} {Band} {Geometric} {Phases} in {One}-{Dimensional} {Systems}},
	volume = {4},
	issn = {2160-3308},
	url = {https://link.aps.org/doi/10.1103/PhysRevX.4.021017},
	doi = {10.1103/PhysRevX.4.021017},
	
	number = {2},
	urldate = {2020-09-29},
	journal = {Physical Review X},
	author = {Xiao, Meng and Zhang, Z. Q. and Chan, C. T.},
	month = apr,
	year = {2014},
	pages = {021017},
}

@article{esmann_topological_2018_1,
	title = {Topological nanophononic states by band inversion},
	volume = {97},
	issn = {2469-9950, 2469-9969},
	url = {https://link.aps.org/doi/10.1103/PhysRevB.97.155422},
	doi = {10.1103/PhysRevB.97.155422},
	
	number = {15},
	urldate = {2020-09-28},
	journal = {Physical Review B},
	author = {Esmann, Martin and Lamberti, Fabrice Roland and Senellart, Pascale and Favero, Ivan and Krebs, Olivier and Lanco, Loïc and Gomez Carbonell, Carmen and Lema{\^i}tre, Aristide and Lanzillotti-Kimura, Norberto Daniel},
	month = apr,
	year = {2018},
	pages = {155422},
}

@article{Zak,
  title = {Berry's phase for energy bands in solids},
  author = {Zak, J.},
  journal = {Phys. Rev. Lett.},
  volume = {62},
  issue = {23},
  pages = {2747--2750},
  numpages = {0},
  year = {1989},
  month = {Jun},
  publisher = {American Physical Society},
  doi = {10.1103/PhysRevLett.62.2747},
  url = {https://link.aps.org/doi/10.1103/PhysRevLett.62.2747}
}

@article{RefInd,
author = {Papatryfonos,Konstantinos  and Angelova,Todora  and Brimont,Antoine  and Reid,Barry  and Guldin,Stefan  and Smith,Peter Raymond  and Tang,Mingchu  and Li,Keshuang  and Seeds,Alwyn J.  and Liu,Huiyun  and Selviah,David R. },
title = {Refractive indices of MBE-grown AlxGa(1-x)As ternary alloys in the transparent wavelength region},
journal = {AIP Advances},
volume = {11},
number = {2},
pages = {025327},
year = {2021},
doi = {10.1063/5.0039631},
URL = {https://doi.org/10.1063/5.0039631},
eprint = {https://doi.org/10.1063/5.0039631}

}

@article{zhao_topological_2018,
	title = {Topological interface modes in local resonant acoustic systems},
	volume = {98},
	url = {https://link.aps.org/doi/10.1103/PhysRevB.98.014110},
	doi = {10.1103/PhysRevB.98.014110},
	number = {1},
	urldate = {2023-05-02},
	journal = {Physical Review B},
	author = {Zhao, Degang and Xiao, Meng and Ling, C. W. and Chan, C. T. and Fung, Kin Hung},
	month = jul,
	year = {2018},
	pages = {014110},
}

@article{zheng_observation_2019,
	title = {Observation of {Edge} {Waves} in a {Two}-{Dimensional} {Su}-{Schrieffer}-{Heeger} {Acoustic} {Network}},
	volume = {12},
	url = {https://link.aps.org/doi/10.1103/PhysRevApplied.12.034014},
	doi = {10.1103/PhysRevApplied.12.034014},
	number = {3},
	urldate = {2023-05-02},
	journal = {Physical Review Applied},
	author = {Zheng, Li-Yang and Achilleos, Vassos and Richoux, Olivier and Theocharis, Georgios and Pagneux, Vincent},
	month = sep,
	year = {2019},
	pages = {034014},
}

@article{delsing_2019_2019,
	title = {The 2019 surface acoustic waves roadmap},
	volume = {52},
	issn = {0022-3727},
	url = {https://iopscience.iop.org/article/10.1088/1361-6463/ab1b04/meta},
	doi = {10.1088/1361-6463/ab1b04},
	number = {35},
	urldate = {2023-05-03},
	journal = {Journal of Physics D: Applied Physics},
	author = {Delsing, Per and Cleland, Andrew N. and Schuetz, Martin J. A. and Knörzer, Johannes and Giedke, Géza and Cirac, J. Ignacio and Srinivasan, Kartik and Wu, Marcelo and Balram, Krishna Coimbatore and Bäuerle, Christopher and Meunier, Tristan and Ford, Christopher J. B. and Santos, Paulo V. and Cerda-Méndez, Edgar and Wang, Hailin and Krenner, Hubert J. and Nysten, Emeline D. S. and Weiß, Matthias and Nash, Geoff R. and Thevenard, Laura and Gourdon, Catherine and Rovillain, Pauline and Marangolo, Max and Duquesne, Jean-Yves and Fischerauer, Gerhard and Ruile, Werner and Reiner, Alexander and Paschke, Ben and Denysenko, Dmytro and Volkmer, Dirk and Wixforth, Achim and Bruus, Henrik and Wiklund, Martin and Reboud, Julien and Cooper, Jonathan M. and Fu, YongQing and Brugger, Manuel S. and Rehfeldt, Florian and Westerhausen, Christoph},
	month = jul,
	year = {2019},
	pages = {353001},
}

@article{mathew_synthetic_2020,
	title = {Synthetic gauge fields for phonon transport in a nano-optomechanical system},
	volume = {15},
	copyright = {2020 The Author(s), under exclusive licence to Springer Nature Limited},
	issn = {1748-3395},
	url = {https://www.nature.com/articles/s41565-019-0630-8},
	doi = {10.1038/s41565-019-0630-8},
	number = {3},
	urldate = {2023-05-03},
	journal = {Nature Nanotechnology},
	author = {Mathew, John P. and Pino, Javier del and Verhagen, Ewold},
	month = mar,
	year = {2020},
	pages = {198--202},
}

@article{parappurath_direct_2020,
	title = {Direct observation of topological edge states in silicon photonic crystals: {Spin}, dispersion, and chiral routing},
	volume = {6},
	shorttitle = {Direct observation of topological edge states in silicon photonic crystals},
	url = {https://www.science.org/doi/full/10.1126/sciadv.aaw4137},
	doi = {10.1126/sciadv.aaw4137},
	number = {10},
	urldate = {2023-05-03},
	journal = {Science Advances},
	author = {Parappurath, Nikhil and Alpeggiani, Filippo and Kuipers, L. and Verhagen, Ewold},
	month = mar,
	year = {2020},
	pages = {eaaw4137},
}

@article{arora_direct_2021,
	title = {Direct quantification of topological protection in symmetry-protected photonic edge states at telecom wavelengths},
	volume = {10},
	copyright = {2021 The Author(s)},
	issn = {2047-7538},
	url = {https://www.nature.com/articles/s41377-020-00458-6},
	doi = {10.1038/s41377-020-00458-6},
	number = {1},
	urldate = {2023-05-03},
	journal = {Light: Science \& Applications},
	author = {Arora, Sonakshi and Bauer, Thomas and Barczyk, René and Verhagen, Ewold and Kuipers, L.},
	month = jan,
	year = {2021},
	pages = {9},
}

@article{malz_topological_2019,
	title = {Topological magnon amplification},
	volume = {10},
	copyright = {2019 The Author(s)},
	issn = {2041-1723},
	url = {https://www.nature.com/articles/s41467-019-11914-2},
	doi = {10.1038/s41467-019-11914-2},
	number = {1},
	urldate = {2023-05-03},
	journal = {Nature Communications},
	author = {Malz, Daniel and Knolle, Johannes and Nunnenkamp, Andreas},
	month = sep,
	year = {2019},
	pages = {3937},
}

@article{Wang_2018,
doi = {10.1088/1555-6611/aae194},
url = {https://dx.doi.org/10.1088/1555-6611/aae194},
year = {2018},
month = {oct},
publisher = {IOP Publishing},
volume = {28},
number = {12},
pages = {126206},
author = {Jun Wang and Yiming Bai and Huiyun Liu and Zhuo Cheng and Mingchu Tang and Siming Chen and Jiang Wu and Konstantinos Papatryfonos and Zizhou Liu and Yongqing Huang and Xiaomin Ren},
title = {Optimization of 1.3 µm InAs/GaAs quantum dot lasers epitaxially grown on silicon: taking the optical loss of metamorphic epilayers into account},
journal = {Laser Physics},
}

@article{lateral_Bragg_Papatry,
    author = {Papatryfonos, Konstantinos and Saladukha, Dzianis and Merghem, Kamel and Joshi, Siddharth and Lelarge, Francois and Bouchoule, Sophie and Kazazis, Dimitrios and Guilet, Stephane and Le Gratiet, Luc and Ochalski, Tomasz J. and Huyet, Guillaume and Martinez, Anthony and Ramdane, Abderrahim},
    title = {Laterally coupled distributed feedback lasers emitting at 2 $\mu$m with quantum dash active region and high-duty-cycle etched semiconductor gratings},
    journal = {Journal of Applied Physics},
    volume = {121},
    number = {5},
    year = {2017},
    month = {02},
    issn = {0021-8979},
    doi = {10.1063/1.4975036},
    url = {https://doi.org/10.1063/1.4975036},
   
}

@article{PhysRevB.97.020102,
  title = {Snowflake phononic topological insulator at the nanoscale},
  author = {Brendel, Christian and Peano, Vittorio and Painter, Oskar and Marquardt, Florian},
  journal = {Phys. Rev. B},
  volume = {97},
  issue = {2},
  pages = {020102(R)},
  numpages = {5},
  year = {2018},
  month = {Jan},
  publisher = {American Physical Society},
  doi = {10.1103/PhysRevB.97.020102},
  url = {https://link.aps.org/doi/10.1103/PhysRevB.97.020102}
}

@article{song_plasmonic_2021,
  title = {Plasmonic Topological Metasurface by Encircling an Exceptional Point},
  author = {Song, Qinghua and Odeh, Mutasem and {Z{\'u}{\~n}iga-P{\'e}rez}, Jes{\'u}s and Kant{\'e}, Boubacar and Genevet, Patrice},
  year = {2021},
  month = sep,
  journal = {Science},
  volume = {373},
  number = {6559},
  pages = {1133--1137},
  issn = {0036-8075, 1095-9203},
  doi = {10.1126/science.abj3179},
  urldate = {2023-05-16}
  }

@article{khanikaev_two-dimensional_2017,
  title = {Two-Dimensional Topological Photonics},
  author = {Khanikaev, Alexander B. and Shvets, Gennady},
  year = {2017},
  month = dec,
  journal = {Nature Photonics},
  volume = {11},
  number = {12},
  pages = {763--773},
  issn = {1749-4885, 1749-4893},
  doi = {10.1038/s41566-017-0048-5},
  urldate = {2023-05-29},
  langid = {english},

}

@article{prodan_topological_2009,
  title = {Topological {{Phonon Modes}} and {{Their Role}} in {{Dynamic Instability}} of {{Microtubules}}},
  author = {Prodan, Emil and Prodan, Camelia},
  year = {2009},
  month = dec,
  journal = {Physical Review Letters},
  volume = {103},
  number = {24},
  pages = {248101},
  issn = {0031-9007, 1079-7114},
  doi = {10.1103/PhysRevLett.103.248101},
  urldate = {2023-05-29},
  langid = {english},

}

@article{susstrunk_observation_2015,
  title = {Observation of Phononic Helical Edge States in a Mechanical Topological Insulator},
  author = {S{\"u}sstrunk, Roman and Huber, Sebastian D.},
  year = {2015},
  month = jul,
  journal = {Science},
  volume = {349},
  number = {6243},
  pages = {47--50},
  issn = {0036-8075, 1095-9203},
  doi = {10.1126/science.aab0239},
  urldate = {2023-05-29},
}

@article{he_acoustic_2016,
  title = {Acoustic Topological Insulator and Robust One-Way Sound Transport},
  author = {He, Cheng and Ni, Xu and Ge, Hao and Sun, Xiao-Chen and Chen, Yan-Bin and Lu, Ming-Hui and Liu, Xiao-Ping and Chen, Yan-Feng},
  year = {2016},
  month = dec,
  journal = {Nature Physics},
  volume = {12},
  number = {12},
  pages = {1124--1129},
  issn = {1745-2473, 1745-2481},
  doi = {10.1038/nphys3867},
  urldate = {2023-05-29},
  langid = {english},

}

@article{huber_topological_2016,
  title = {Topological Mechanics},
  author = {Huber, Sebastian D.},
  year = {2016},
  month = jul,
  journal = {Nature Physics},
  volume = {12},
  number = {7},
  pages = {621--623},
  issn = {1745-2473, 1745-2481},
  doi = {10.1038/nphys3801},
  urldate = {2023-05-29},
  langid = {english},

}

@article{priya_perspectives_2023,
  title = {Perspectives on High-Frequency Nanomechanics, Nanoacoustics, and Nanophononics},
  author = {{Priya} and {Cardozo de Oliveira}, E. R. and {Lanzillotti-Kimura}, N. D.},
  year = {2023},
  month = apr,
  journal = {Applied Physics Letters},
  volume = {122},
  number = {14},
  pages = {140501},
  issn = {0003-6951, 1077-3118},
  doi = {10.1063/5.0142925},
  urldate = {2023-05-29},
}

@article{kuznetsov_electrically_2021-1,
  title = {Electrically {{Driven Microcavity Exciton-Polariton Optomechanics}} at 20 {{GHz}}},
  author = {Kuznetsov, Alexander S. and Machado, Diego H. O. and Biermann, Klaus and Santos, Paulo V.},
  year = {2021},
  month = apr,
  journal = {Physical Review X},
  volume = {11},
  number = {2},
  pages = {021020},
  issn = {2160-3308},
  doi = {10.1103/PhysRevX.11.021020},
  urldate = {2023-08-14},
  langid = {english},

}

@article{torres_giant_2019,
  title = {Giant {{Electrophononic Response}} in {{PbTiO}} 3 by {{Strain Engineering}}},
  author = {Torres, Pol and Iniguez, Jorge and Rurali, Riccardo},
  year = {2019},
  month = oct,
  journal = {Physical Review Letters},
  volume = {123},
  number = {18},
  pages = {185901},
  issn = {0031-9007, 1079-7114},
  doi = {10.1103/PhysRevLett.123.185901},
  urldate = {2023-08-14}
}

@article{li_remarkably_2021,
  title = {Remarkably {{Weak Anisotropy}} in {{Thermal Conductivity}} of {{Two-Dimensional Hybrid Perovskite Butylammonium Lead Iodide Crystals}}},
  author = {Li, Chen and Ma, Hao and Li, Tianyang and Dai, Jinghang and Rasel, Md. Abu Jafar and Mattoni, Alessandro and Alatas, Ahmet and Thomas, Malcolm G. and Rouse, Zachary W. and Shragai, Avi and Baker, Shefford P. and Ramshaw, B. J. and Feser, Joseph P. and Mitzi, David B. and Tian, Zhiting},
  year = {2021},
  month = may,
  journal = {Nano Letters},
  volume = {21},
  number = {9},
  pages = {3708--3714},
  publisher = {{American Chemical Society}},
  issn = {1530-6984},
  doi = {10.1021/acs.nanolett.0c04550},
  urldate = {2023-08-14},
}

@article{bencivenga_nanoscale_2019,
  title = {Nanoscale Transient Gratings Excited and Probed by Extreme Ultraviolet Femtosecond Pulses},
  author = {Bencivenga, F. and Mincigrucci, R. and Capotondi, F. and Foglia, L. and Naumenko, D. and Maznev, A. A. and Pedersoli, E. and Simoncig, A. and Caporaletti, F. and Chiloyan, V. and Cucini, R. and Dallari, F. and Duncan, R. A. and Frazer, T. D. and Gaio, G. and Gessini, A. and Giannessi, L. and Huberman, S. and Kapteyn, H. and Knobloch, J. and Kurdi, G. and Mahne, N. and Manfredda, M. and Martinelli, A. and Murnane, M. and Principi, E. and Raimondi, L. and Spampinati, S. and Spezzani, C. and Trov{\`o}, M. and Zangrando, M. and Chen, G. and Monaco, G. and Nelson, K. A. and Masciovecchio, C.},
  year = {2019},
  month = jul,
  journal = {Science Advances},
  volume = {5},
  number = {7},
  pages = {eaaw5805},
  issn = {2375-2548},
  doi = {10.1126/sciadv.aaw5805},
  urldate = {2023-08-14},
}

@article{buhler_-chip_2022,
  title = {On-Chip Generation and Dynamic Piezo-Optomechanical Rotation of Single Photons},
  author = {B{\"u}hler, Dominik D. and Wei{\ss}, Matthias and {Crespo-Poveda}, Antonio and Nysten, Emeline D. S. and Finley, Jonathan J. and M{\"u}ller, Kai and Santos, Paulo V. and {de Lima}, Mauricio M. and Krenner, Hubert J.},
  year = {2022},
  month = nov,
  journal = {Nature Communications},
  volume = {13},
  number = {1},
  pages = {6998},
  issn = {2041-1723},
  doi = {10.1038/s41467-022-34372-9},
  urldate = {2023-08-14},
  
}

@article{meng_measurement-based_2022,
  title = {Measurement-Based Preparation of Multimode Mechanical States},
  author = {Meng, Chao and Brawley, George A. and Khademi, Soroush and Bridge, Elizabeth M. and Bennett, James S. and Bowen, Warwick P.},
  year = {2022},
  month = may,
  journal = {Science Advances},
  volume = {8},
  number = {21},
  pages = {eabm7585},
  publisher = {{American Association for the Advancement of Science}},
  doi = {10.1126/sciadv.abm7585},
  urldate = {2023-08-14},
}

@article{mahboob_phonon-cavity_2012,
  title = {Phonon-Cavity Electromechanics},
  author = {Mahboob, I. and Nishiguchi, K. and Okamoto, H. and Yamaguchi, H.},
  year = {2012},
  month = may,
  journal = {Nature Physics},
  volume = {8},
  number = {5},
  pages = {387--392},
  publisher = {{Nature Publishing Group}},
  issn = {1745-2481},
  doi = {10.1038/nphys2277},
  urldate = {2023-08-14},
  copyright = {2012 Springer Nature Limited},
  langid = {english},
}

@article{rodriguez_2023,
	title = {Topological nanophononic interface states using high-order bandgaps in the one-dimensional {Su}-{Schrieffer}-{Heeger} model},
	volume = {108},
	issn = {2469-9950, 2469-9969},
	url = {https://link.aps.org/doi/10.1103/PhysRevB.108.205301},
	doi = {10.1103/PhysRevB.108.205301},
	number = {20},
	urldate = {2025-12-07},
	journal = {Physical Review B},
	author = {Rodriguez, A. and Papatryfonos, K. and Cardozo De Oliveira, E. R. and Lanzillotti-Kimura, N. D.},
	month = nov,
	year = {2023},
	pages = {205301},
}

@inproceedings{papatryfonos_SPIE_2024,
	address = {Strasbourg, France},
	title = {High-order topological states using alignment of different bandgaps in {1D} superlattices},
	isbn = {978-1-5106-7300-7 978-1-5106-7301-4},
	url = {https://www.spiedigitallibrary.org/conference-proceedings-of-spie/12991/3017060/High-order-topological-states-using-alignment-of-different-bandgaps-in/10.1117/12.3017060.full},
	doi = {10.1117/12.3017060},
	urldate = {2025-12-07},
	booktitle = {Nanophotonics {X}},
	publisher = {SPIE},
	author = {Papatryfonos, Konstantinos and Rodriguez, Anne and Cardozo De Oliveira, Edson R. and Lanzillotti-Kimura, Norberto Daniel},
	editor = {Andrews, David L. and Bain, Angus J. and Ambrosio, Antonio},
	month = jun,
	year = {2024},
	pages = {25},
}

@misc{Chushuang_2024,
	title = {Interference of ultrahigh frequency acoustic phonons from distant quasi-continuous sources},
	url = {http://arxiv.org/abs/2407.06821},
	doi = {10.48550/arXiv.2407.06821},
	urldate = {2025-12-07},
	publisher = {arXiv},
	author = {Xiang, C. and Oliveira, E. R. Cardozo de and Sandeep, S. and Papatryfonos, K. and Morassi, M. and Gratiet, L. Le and Harouri, A. and Sagnes, I. and Lemaitre, A. and Ortiz, O. and Esmann, M. and Lanzillotti-Kimura, N. D.},
	month = jul,
	year = {2024},
	note = {arXiv:2407.06821 [cond-mat]},
}

@article{papatryfonos_2025,
	title = {Low‐{Defect} {Quantum} {Dot} {Lasers} {Directly} {Grown} on {Silicon} {Exhibiting} {Low} {Threshold} {Current} and {High} {Output} {Power} at {Elevated} {Temperatures}},
	volume = {6},
	copyright = {http://creativecommons.org/licenses/by/4.0/},
	issn = {2699-9293, 2699-9293},
	url = {https://advanced.onlinelibrary.wiley.com/doi/10.1002/adpr.202400082},
	doi = {10.1002/adpr.202400082},
	number = {3},
	urldate = {2025-12-07},
	journal = {Advanced Photonics Research},
	author = {Papatryfonos, Konstantinos and Girard, Jean‐Christophe and Tang, Mingchu and Deng, Huiwen and Seeds, Alwyn J. and David, Christophe and Rodary, Guillemin and Liu, Huiyun and Selviah, David R.},
	month = mar,
	year = {2025},
	pages = {2400082},
}

@article{schmidt_coupled_2021,
	title = {Coupled topological interface states},
	volume = {103},
	issn = {2469-9950, 2469-9969},
	url = {https://link.aps.org/doi/10.1103/PhysRevB.103.085412},
	doi = {10.1103/PhysRevB.103.085412},
	number = {8},
	urldate = {2025-12-07},
	journal = {Physical Review B},
	author = {Schmidt, Christoph and Palatnik, Alexander and Sudzius, Markas and Meister, Stefan and Leo, Karl},
	month = feb,
	year = {2021},
	pages = {085412},
}

@article{sharma_2023,
	title = {Coupling of topological interface states in {1D} photonic crystal},
	volume = {137},
	issn = {09253467},
	url = {https://linkinghub.elsevier.com/retrieve/pii/S0925346723000794},
	doi = {10.1016/j.optmat.2023.113508},
	urldate = {2025-12-07},
	journal = {Optical Materials},
	author = {Sharma, Richa and Jena, Shuvendu and Udupa, Dinesh V.},
	month = mar,
	year = {2023},
	pages = {113508},

}

@article{mlayahRamanBrillouinLightScattering2007,
  title = {Raman-{{Brillouin}} Light Scattering in Low-Dimensional Systems: {{Photoelastic}} Model versus Quantum Model},
  shorttitle = {Raman-{{Brillouin}} Light Scattering in Low-Dimensional Systems},
  author = {Mlayah, Adnen and Huntzinger, Jean-Roch and Large, Nicolas},
  year = {2007},
  month = jun,
  journal = {Physical Review B},
  volume = {75},
  pages = {245303},
  doi = {10.1103/PhysRevB.75.245303},
  copyright = {http://link.aps.org/licenses/aps-default-license}
}

@article{groenenInelasticLightScattering2008,
  title = {Inelastic Light Scattering by Longitudinal Acoustic Phonons in Thin Silicon Layers: {{From}} Membranes to Silicon-on-Insulator Structures},
  shorttitle = {Inelastic Light Scattering by Longitudinal Acoustic Phonons in Thin Silicon Layers},
  author = {Groenen, J. and Poinsotte, F. and Zwick, A. and Sotomayor Torres, C. M. and Prunnila, M. and Ahopelto, J.},
  year = {2008},
  month = jan,
  journal = {Physical Review B},
  volume = {77},
  pages = {045420},
  doi = {10.1103/PhysRevB.77.045420},
  copyright = {http://link.aps.org/licenses/aps-default-license}
}

@article{jusserandSelectiveResonantInteraction2013,
  title = {Selective Resonant Interaction between Confined Excitons and Folded Acoustic Phonons in {{GaAs}}/{{AlAs}} Multi-Quantum Wells},
  author = {Jusserand, B.},
  year = {2013},
  month = aug,
  journal = {Applied Physics Letters},
  volume = {103},
  pages = {093112},
  doi = {10.1063/1.4817647}
}

@article{jusserandPolaritonResonancesUltrastrong2015,
  title = {Polariton {{Resonances}} for {{Ultrastrong Coupling Cavity Optomechanics}} in {{GaAs}} / {{AlAs Multiple Quantum Wells}}},
  author = {Jusserand, B. and Poddubny, A. N. and Poshakinskiy, A. V. and Fainstein, A. and Lemaitre, A.},
  year = {2015},
  month = dec,
  journal = {Physical Review Letters},
  volume = {115},
  pages = {267402},
  doi = {10.1103/PhysRevLett.115.267402},
  copyright = {http://link.aps.org/licenses/aps-default-license}
}

@article{PascualWinterSpectralResponsesPhonon2012,
  title = {Spectral Responses of Phonon Optical Generation and Detection in Superlattices},
  author = {{Pascual-Winter}, M. F. and Fainstein, A. and Jusserand, B. and Perrin, B. and Lema{\^i}tre, A.},
  year = {2012},
  month = jun,
  journal = {Physical Review B},
  volume = {85},
  pages = {235443},
  doi = {10.1103/PhysRevB.85.235443},
  copyright = {http://link.aps.org/licenses/aps-default-license}
}

@incollection{fainsteinRamanScatteringResonant2006,
  title = {Raman {{Scattering}} in {{Resonant Cavities}}},
  booktitle = {Light {{Scattering}} in {{Solid IX}}},
  author = {Fainstein, Alejandro and Jusserand, Bernard},
  editor = {Cardona, Manuel and Merlin, Roberto},
  year = {2006},
  volume = {108},
  pages = {17--110},
  publisher = {Springer Berlin Heidelberg},
  address = {Berlin, Heidelberg},
  doi = {10.1007/978-3-540-34436-0_2},
  isbn = {978-3-540-34435-3 978-3-540-34436-0}
}

@article{leeElasticTopologicalInterface2022,
  title = {Elastic Topological Interface States and Voltage Feeder by Breaking Inversion Symmetry on Thin Plates},
  author = {Lee, Dongwoo and {Lanzillotti-Kimura}, N. D. and Li, Jensen and Rho, Junsuk},
  year = {2022},
  month = sep,
  journal = {Physical Review B},
  volume = {106},
  pages = {104107},
  doi = {10.1103/PhysRevB.106.104107}
}

@article{fainsteinStrongOpticalMechanicalCoupling2013,
  title = {Strong {{Optical-Mechanical Coupling}} in a {{Vertical GaAs}}/{{AlAs Microcavity}} for {{Subterahertz Phonons}} and {{Near-Infrared Light}}},
  author = {Fainstein, A. and {Lanzillotti-Kimura}, N. D. and Jusserand, B. and Perrin, B.},
  year = {2013},
  month = jan,
  journal = {Physical Review Letters},
  volume = {110},
  pages = {037403},
  doi = {10.1103/PhysRevLett.110.037403},
  copyright = {http://link.aps.org/licenses/aps-default-license}
}

@article{tamuraAcousticphononTransmissionQuasiperiodic1987,
  title = {Acoustic-Phonon Transmission in Quasiperiodic Superlattices},
  author = {Tamura, S. and Wolfe, J. P.},
  year = {1987},
  month = aug,
  journal = {Physical Review B},
  volume = {36},
  pages = {3491--3494},
  doi = {10.1103/PhysRevB.36.3491},
  copyright = {http://link.aps.org/licenses/aps-default-license}
}

\end{document}